\documentclass[dvipsnames]{article} %
\usepackage{colm2024_conference}

\usepackage{titletoc}
\usepackage{titlesec}
\usepackage{booktabs}
\usepackage{graphicx}
\usepackage{enumitem}
\usepackage{wrapfig}
\usepackage{algorithm}
\usepackage{algpseudocode}
\usepackage{pifont}
\usepackage{tikz}
\usetikzlibrary{fadings}
\usepackage{microtype}
\usepackage{amsmath}
\usepackage{colortbl}
\usepackage[utf8]{inputenc}
\definecolor{lightgray}{rgb}{0.9,0.9,0.9}
\usepackage{caption}
\usepackage[utf8]{inputenc}
\usepackage[T1]{fontenc}
\usepackage{hyperref}
\usepackage{url}
\usepackage{booktabs}
\usepackage{amsfonts}
\usepackage{amsmath}
\usepackage{amssymb}
\usepackage{nicefrac}
\usepackage{microtype}
\usepackage{xcolor}

\usepackage{graphicx}
\usepackage{multirow}
\usepackage{array}
\usepackage{tabularx}
\usepackage{makecell}
\usepackage{subcaption}
\usepackage{wrapfig}
\usepackage{enumitem}
\usepackage{colortbl}
\usepackage{adjustbox}
\usepackage{pifont}
\usepackage{xspace}
\usepackage{subcaption}
\usepackage{arydshln}
\usepackage{CJKutf8}
\usepackage{tcolorbox}
\tcbuselibrary{skins,breakable}

\definecolor{giraffeyellow}{RGB}{255,185,15}       % 框架/标题背景：长颈鹿金黄
\definecolor{giraffebg}{RGB}{255,250,235}          % 内容区背景：淡奶黄
\definecolor{giraffeframe}{RGB}{230,165,0}         % 边框线：略深的琥珀黄
\definecolor{giraffetext}{RGB}{140,85,0}           % 文字标签：深棕黄（可读性好）

\tcbset{
  medprompt/.style={
    enhanced, breakable,
    colback=giraffebg, colframe=giraffeframe,
    coltitle=white, fonttitle=\bfseries\small,
    colbacktitle=giraffeyellow,
    boxrule=0.5pt, arc=2pt,
    left=6pt, right=6pt, top=4pt, bottom=4pt,
  }
}

\usepackage{xcolor}

\definecolor{e2ecolor}{RGB}{178, 76, 76}      % soft academic red
\definecolor{cascolor}{RGB}{184, 134, 38}     % muted gold
\definecolor{reascolor}{RGB}{67, 112, 164}    % steel blue

\definecolor{e2ebg}{RGB}{251, 240, 240}       % very light red
\definecolor{casbg}{RGB}{252, 246, 231}       % very light gold
\definecolor{reasbg}{RGB}{238, 244, 250}      % very light blue

\newcommand{\bestE}[1]{\scalebox{1.1}{\textbf{\textcolor{e2ecolor}{#1}}}}
\newcommand{\bestC}[1]{\scalebox{1.1}{\textbf{\textcolor{cascolor}{#1}}}}
\newcommand{\bestR}[1]{\scalebox{1.1}{\textbf{\textcolor{reascolor}{#1}}}}

\definecolor{mygray}{gray}{.92}
\definecolor{ForestGreen}{RGB}{34,139,34}
\definecolor{Forestred}{RGB}{220,50,50}

\newcommand{\CHECK}{\textcolor{ForestGreen}{\ding{52}}}
\newcommand{\CROSS}{\textcolor{Forestred}{\ding{55}}}

\newcommand{\benchname}{\textsc{VoiceGiraffe}\xspace}

\title{\benchname: A Benchmark for Extreme Long-Context Audio-Language Understanding}

\author{%
  Jashin Ye\thanks{Equal contribution.},\quad Dongxiao Wang\footnotemark[1],\quad  Yixuan Ye,\quad  Sashuai Zhou,\quad  Weihuang Lin,\\[1pt] 
  Mingyang Han,\quad  Kunpeng Wang,\quad  Zeyu Yuan,\quad  Boyu Li,\quad  Haoxiang Shi,\\[1pt]
  Jingchen Shu,\quad  Jun Song\thanks{Corresponding author.},\quad  Bo Zheng \\[3pt]
  \textbf{Future Living Lab, Alibaba} \\[3pt]
}

\begin{document}
\begin{CJK*}{UTF8}{gbsn}

\maketitle

\begin{abstract}
% \vspace{-2mm}
While large audio language models (LALMs) have achieved remarkable progress in audio processing at the second- or minute-level scale, understanding hour-level audio remains a fundamental bottleneck. Existing benchmarks predominantly rely on short clips or artificially concatenated segments, failing to faithfully assess LALMs' capacity for long-range information comprehension in real-world scenarios such as podcasts and lengthy speeches. To address this gap, we introduce \benchname, a novel benchmark designed to rigorously evaluate LALMs across diverse real-world scenarios, modalities, and languages under long-context settings.  It comprises 1,500 curated triplets structured into a dual-level taxonomy of single-hop perception and multi-hop reasoning. 
Single-hop questions assess temporal grounding, semantics, paralinguistics, and acoustic events, while multi-hop questions require models to aggregate evidence across multiple non-contiguous segments. 
We evaluate a broad suite of open-source and proprietary LALMs against human performance. Results underscore three fundamental findings. First, \benchname remains highly challenging and far from saturation, with only one end-to-end (E2E) LALM surpassing the human reference, and no open-source model reaches the passing score even under cascaded caption aggregation.
Second, we show that no single inference paradigm universally dominates. The E2E inference benefits models with native long-context audio understanding, cascaded caption aggregation stabilizes small models overwhelmed by hour-scale audio, and reasoning-enhanced cascading with external LLM helps weaker models but can bottleneck stronger proprietary systems.
Third, we reveal long-range memory persistence as a key bottleneck. LALMs are better at answering questions that require connecting salient causal cues than those requiring sustained tracking of sparse events across long audio, whereas humans show the opposite pattern. This suggests that current LALMs can reason over prominent evidence once it is localized, but struggle to continuously memorize and retrieve event states over hour-scale contexts. 
These findings position \benchname as a challenging and diagnostic testbed for long-form audio understanding, highlighting the need for LALMs with persistent memory and robust long-range aggregation. 
%Our data and evaluation code are available on \href{https://github.com/VoiceGiraffe/VoiceGiraffe}{https://github.com/VoiceGiraffe/VoiceGiraffe}

% We hope \benchname will catalyse future advances in this important yet underexplored area.

\end{abstract}

% \vspace{-0.4cm}
\section{Introduction}
\label{sec:introduction}
% \vspace{-0.2cm}

\begin{figure}[t]
    \centering
    \includegraphics[width=1.0\linewidth]{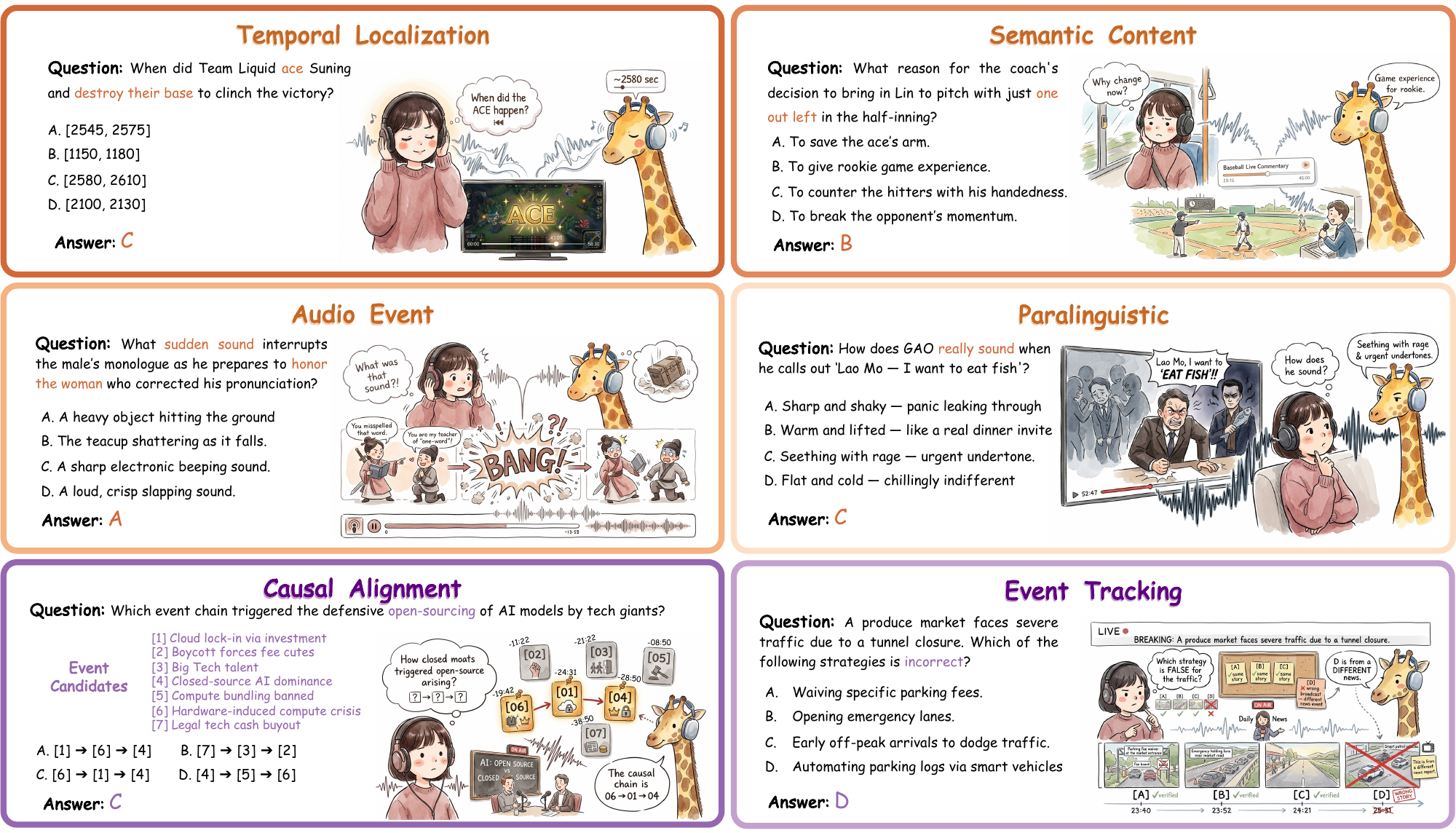}
    \vspace{-3.5mm}
    \caption{Examples of \benchname{} across five real-world domains: e-sports commentary, sports broadcast, TV drama, news, and podcast. 
    \textcolor{orange}{\textit{Orange}} panels show single-hop questions (temporal localization, semantics, audio events, paralinguistics). \textcolor{violet}{\textit{Purple}} panels show multi-hop questions (causal reasoning and event tracking) requiring evidence aggregation across long spans.}
    \label{fig:teaser}
    % \vspace{-0.52cm}
\end{figure}

Large Audio Language Models (LALMs)~\cite{qwen-audio, qwen2-audio, audio-flamingo, audio-flamingo2, salmonn-omni} have rapidly emerged as a unifying paradigm for speech, sound and music understanding, combining an audio encoder with a large language model to address diverse auditory tasks such as automatic speech recognition, audio captioning and audio question answering~\cite{mmau,airbench,mmar}. 
As LALMs are increasingly deployed as audio assistants for meetings, podcasts, live broadcasts, and long interviews, reasoning over \emph{hour-scale} audio inputs has become a crucial capability.
However, current LALMs face fundamental challenges in long-context understanding and reasoning, stemming from the high audio token rate and insufficient pre-training on long-horizon audio understanding. 
The computational overhead imposed by dense tokens, coupled with limited exposure to longer contexts during training, systematically impedes effective long-horizon scaling. 
Consequently, models exhibit significant performance degradation when applied to long-form sequences. 

A critical reason the community has struggled to tackle this challenge is that evaluation protocols lag far behind both model capabilities and real-world requirements. 
As summarized in Table~\ref{tab:compare_prior_a}, existing short-form benchmarks such as MMAU~\cite{mmau}, MMAR~\cite{mmar}, AudioBench~\cite{audiobench}, and AIR-Bench~\cite{airbench} provide robust and comprehensive evaluation of audio understanding and reasoning for clips of 10–30 seconds, yet severe limitations emerge at longer temporal scales. 
While recent efforts have extended evaluation to minute-level contexts, such as AudioMarathon~\cite{audiomarathon}, ChronosAudio~\cite{chronosaudio}, and LongSpeech~\cite{longspeech}, these benchmarks still fall short of real-world scenarios requiring hour-level fine-grained understanding, such as sports events, TV dramas, and podcasts. 
More importantly, several benchmarks construct long-context inputs by concatenating short audio clips rather than using native long-form recordings, making it difficult to capture the natural temporal continuity and sparse long-range dependencies present in real-world audio. 
Beyond the duration and authenticity gap, existing benchmarks tend to focus on single aspects and lack comprehensive integration of bilingual coverage, cross-domain diversity, and multi-hop reasoning.

To address these gaps, we introduce \benchname, the first extremely long-context audio question-answer (AQA) benchmark for evaluating long-context understanding capacities in LALMs. 
As illustrated in Figure~\ref{fig:teaser} and Table~\ref{tab:compare_prior}, \benchname is built on three pillars. \textbf{First, hour-scale context}. \benchname collects 123 long-form recordings that total approximately 113.1 hours, with an average duration of 55.2 minutes and 34\% of recordings exceeding one hour. To our knowledge, it is the first open-domain AQA benchmark at this scale. \textbf{Second, bilingual, open-domain, and full-modality coverage.}  The benchmark includes five domains across bilingual: sports, e-sports commentary, TV dramas, news, and podcasts. All recordings contain natural mixtures of speech, sound effects, and background music, challenging models on both verbal comprehension and cross-modal reasoning. \textbf{Third, a two-tier task taxonomy from single-hop perception to multi-hop reasoning.} Tier-1 questions assess single-hop perception across four categories: temporal localization, semantic content, audio event detection, and paralinguistic analysis. Tier-2 questions target multi-hop reasoning through causal event tracing and state-trajectory modeling, requiring evidence synthesis from 2 to 4 non-contiguous segments spanning tens of minutes.

Based on \benchname, we evaluate 9 open-source and 4 proprietary audio-capable models together with a human reference.
Our analysis reveals four key findings. 
First, \benchname presents substantial challenges to current LALMs. Among the models capable of hour-scale E2E inference, only Qwen3.5-Omni-Plus surpasses the human reference, while open-source systems remain below the passing threshold under cascaded caption aggregation without any enhancement. 
Second, the optimal inference paradigm is highly model-dependent.
E2E inference can better preserve fine-grained perceptual cues for models with strong native long-context audio understanding ability, whereas cascaded caption aggregation provides a more stable way for models overwhelmed by long-form audio. 
Reasoning-enhanced cascading further improves open-source models by compensating for weak long-range aggregation, but can also bottleneck strong proprietary systems when the external reasoning model is less capable than the evaluated model itself.
Third, long-term memory persistence remains a key bottleneck.
Unlike humans, who are better at tracking sparse states over extended context, LALMs are stronger at reasoning over salient causal cues but struggle to memorize and retrieve states across distant clips.

In summary, our contributions are listed as follows:

\begin{itemize}[leftmargin=14pt, topsep=0pt, itemsep=2pt,
partopsep=0pt, parsep=0pt]
    \item We present \benchname, the first bilingual, hour-scale audio benchmark for LALMs, covering five real-world scenarios with naturally interleaved speech, sound and music.
    \item We design a two-tier task taxonomy with 1,500 carefully designed AQA pairs that progress from perception to reasoning, supporting systematic evaluation of long-context understanding.
    \item We conduct comprehensive analysis and comparison experiments across broad suite of LALMs together with a human reference, identifying long-context degradation, long-form memory failures and language bias as the key bottlenecks of current models. 
\end{itemize}

% \vspace{-0.1cm}
\section{Related Work}
\label{sec:related_works}
\vspace{-0.1cm}

\subsection{Large Audio-Language Models}
\vspace{-0.05cm}

The landscape of Large Audio Language Models (LALMs) has expanded rapidly along both open-source and proprietary tracks.
On the open-source side, the Qwen-Audio series~\citep{qwen-audio, qwen2-audio} pioneered general audio understanding, later extended by Qwen2.5-Omni and Qwen3-Omni~\citep{qwen2omni,qwen3omni} to unified audio--vision--language reasoning.
Audio Flamingo~\citep{audio-flamingo} and its successor Audio Flamingo~2~\citep{audio-flamingo2} introduced few-shot and long-audio capabilities, while SALMONN~\citep{salmonn}, GAMA~\citep{gama}, Baichuan-Omni~\citep{baichuan-omni}, Phi-4-Multimodal~\citep{phi4}, and MiMo-Audio~\citep{mimo-audio} contribute diverse architectural designs audio encoders, mixture-of-experts, and end-to-end streaming.
Among proprietary systems, GPT-4o-Audio~\citep{gpt-4o} and the Gemini~2.5 series family~\citep{gemini} demonstrate strong integrated multimodal reasoning, and Qwen3.5-Omni-PLUS now supports a 256K-token context window~\citep{qwen3.5omni}.
Despite these ever-growing context capacities and reasoning abilities, \emph{no systematic evaluation protocol exists for hour-scale naturalistic audio}, a setting that is increasingly common in real-world applications such as podcasts, meetings, live broadcasts, and long interviews. 
This gap motivates us to establish a rigorous evaluation framework for hour-scale audio understanding.

% Table: dataset_comparison.tex
% Comparison of Odyssey with prior audio benchmarks + key statistics (side-by-side).
\begin{table}[t]
\centering
\caption{(\textbf{a}) Comparison of \benchname with representative audio benchmarks across four key dimensions: average per-sample duration, heterogeneous modality coverage, bilingual contexts, and multi-hop reasoning. (\textbf{b}) Preliminary statistics of \benchname.}
\vspace{-1.6mm}
\label{tab:compare_prior}
\begin{subtable}[t]{0.68\linewidth}
\centering
\caption{\textit{Benchmark Comparison}}
\label{tab:compare_prior_a}
\vspace{-1.2mm}
\resizebox{\linewidth}{!}{
\centering
\begin{tabular}{lcccccc}
\toprule
\multirow{2.5}{*}{\textbf{Benchmark}}
& \multirow{2.5}{*}{\textbf{Duration}}
& \multicolumn{3}{c}{\textbf{Modality}}
& \multirow{2.5}{*}{\textbf{Bilingual}}
& \multirow{2.5}{*}{\textbf{Multi-Hop}} \\
\cmidrule(lr){3-5} & & \textbf{Speech} & \textbf{Sound} & \textbf{Music} & & \\
\midrule
MMAU~\cite{mmau}                         & $\sim$10\,sec   & \CHECK & \CHECK & \CHECK & \CROSS & \CROSS \\
MMAR~\cite{mmar}                         & $\sim$20\,sec     & \CHECK & \CHECK & \CHECK & \CROSS & \CHECK  \\
AudioBench~\cite{audiobench}             & $\sim$14\,sec    & \CHECK & \CHECK & \CHECK & \CROSS & \CROSS \\
AIR-Bench~\cite{airbench}                & $\sim$35\,sec   & \CHECK & \CHECK & \CHECK & \CROSS & \CROSS \\
AudioMarathon~\cite{audiomarathon}       & $\sim$4\,min  & \CHECK & \CHECK & \CHECK & \CROSS & \CHECK  \\
LongSpeech~\cite{longspeech}             & $\sim$10\,min    & \CHECK & \CROSS & \CROSS & \CHECK & \CROSS \\
ChronosAudio~\cite{chronosaudio}         & $\sim$5\,min   & \CHECK & \CROSS & \CROSS & \CROSS & \CROSS \\
BLAB~\cite{blab}                         & $\sim$51\,min   & \CHECK & \CROSS & \CROSS & \CROSS & \CROSS \\
\midrule
\textbf{\benchname{}}~\textit{(ours)}
                                         & \textbf{$\sim$1\,hr} & \CHECK & \CHECK & \CHECK & \CHECK & \CHECK \\
\bottomrule
\end{tabular}}
\end{subtable}%
\hfill
\begin{subtable}[t]{0.29\linewidth}
\caption{\textit{Benchmark Statistics}}
\label{tab:compare_prior_b}
\vspace{-1.3mm}
\centering
\resizebox{\linewidth}{!}{
\begin{tabular}{lc}
\toprule
\textbf{Statistics} & \textbf{Value} \\
\midrule
Total recordings          & 123 \\
Total audio duration (h)  & 113.1 \\
Avg.\ duration (min)      & 55.2 \\
Sub-tasks               & 6 \\
Data-domains               & 5 \\
Languages                 & EN / ZH \\
\midrule
Total QA items            & 1,500 \\
Single-hop items       & 1,000 \\
Multi-hop items        & 500 \\
\bottomrule
\end{tabular}}
\end{subtable}

\vspace{0mm}
\end{table}

\subsection{Audio Understanding Benchmarks}
\vspace{-0.05cm}

\paragraph{Second-Level Benchmarks.}
AudioBench~\citep{audiobench}, AIR-Bench~\citep{airbench}, MMAU~\citep{mmau}, and MMAR~\citep{mmar} evaluate LALMs on clips of 10--30 seconds, providing comprehensive coverage of perception, instruction following, and reasoning within a single auditory scene. While rigorous in scope, short-context evaluation fundamentally limits assessment of two critical long-context capabilities: (\emph{i}) maintaining memory of important information across extended listening spans, and (\emph{ii}) aggregating evidence from multiple non-contiguous segments separated by minutes or hours. In contrast, we bridge this gap with \benchname designed to probe these capabilities in the extreme long-context regime.

\paragraph{Minute-Level Benchmarks.}
Recent benchmarks have begun extending evaluation timescales into the minute regime, with efforts spanning 3--10 minutes: AudioMarathon~\citep{audiomarathon}, LongSpeech~\citep{longspeech}, ChronosAudio~\citep{chronosaudio}.
While these efforts mark valuable progress, they fall short of addressing the real-world challenge users increasingly face, namely rapidly extracting information from hour-long podcasts, lectures, and sports commentary. 
More importantly, several benchmarks construct long-context inputs by concatenating short audio clips rather than using native long-form recordings. 
Even BLAB~\citep{blab}, which extends to 51 minutes, remains confined to speech-specific evaluation, without testing multi-domain acoustic understanding or complex temporal reasoning over hour-scale inputs. 
To bridge this gap, our benchmark targets the hour scale with multi-domain coverage and questions requiring multi-hop aggregation.

As summarised in Table~\ref{tab:compare_prior}, \benchname addresses three orthogonal limitations simultaneously:
(\textbf{\emph{i}})~it evaluates on \textbf{hour-scale naturalistic audio} drawn from intact real-world recordings rather than synthetic concatenations;
(\textbf{\emph{ii}})~it covers \textbf{bilingual} (English + Chinese) content across \textbf{five domains} with naturally interleaved speech, music, and environmental sound;
and (\textbf{\emph{iii}})~it introduces a \textbf{perception-to-reasoning} two-tier ontology for comprehensive evaluation across long-form contexts.

\section{\benchname}
\label{sec:benchmark}

\subsection{Overview}
\label{sec:benchmark_overview}
% \vspace{-0.15cm}

\begin{figure}[h]
  \centering
  \begin{subfigure}[b]{0.32\textwidth}
    \centering
    \includegraphics[width=0.9\linewidth]{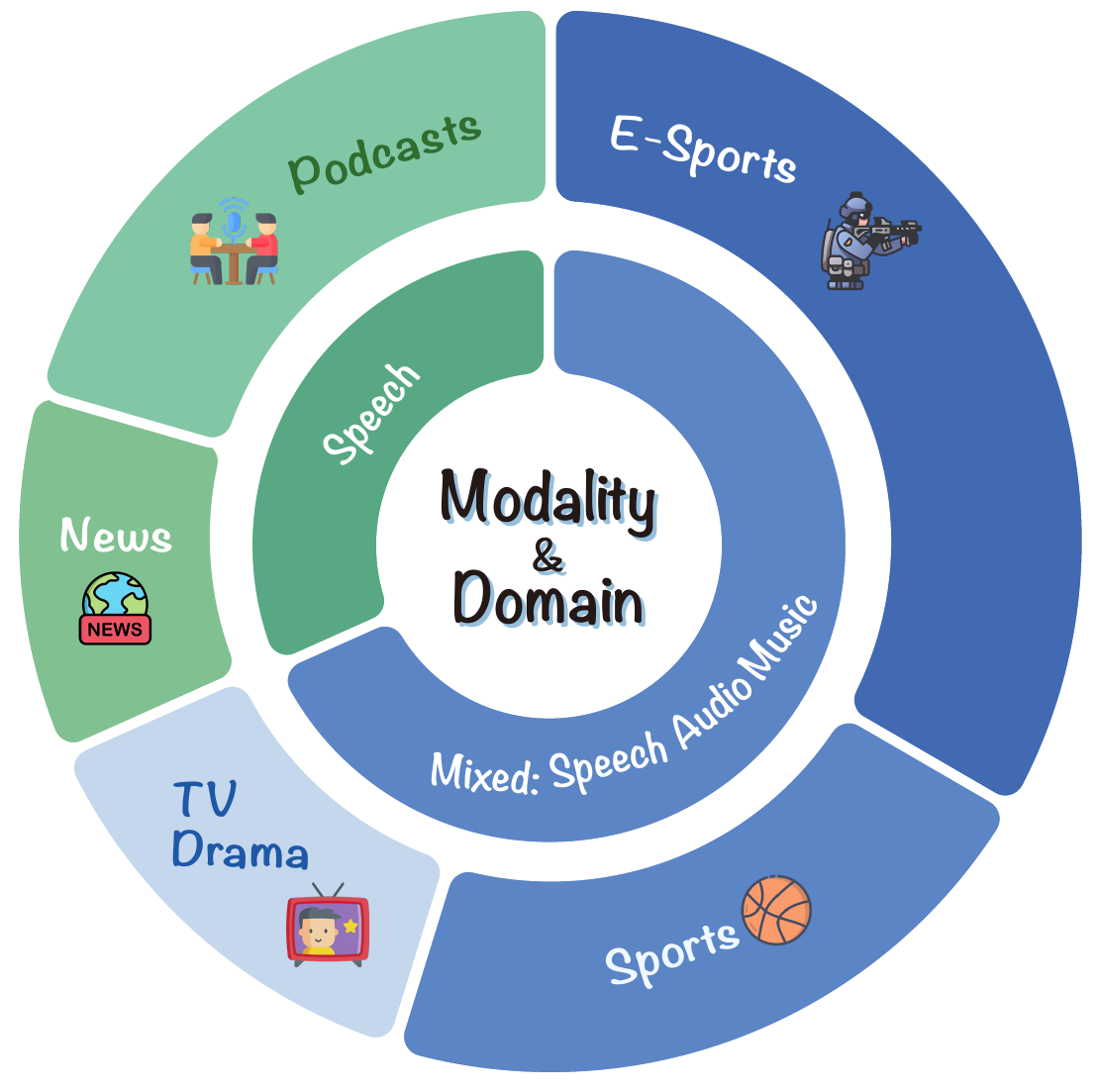}
    % \fbox{\rule[-.5cm]{0cm}{3.0cm} \rule[-.5cm]{4.2cm}{0cm}}
    \caption{Data Distribution}
    \label{fig:modality_pie}
  \end{subfigure}\hfill
  \begin{subfigure}[b]{0.32\textwidth}
    \centering
    \includegraphics[width=0.9\linewidth]{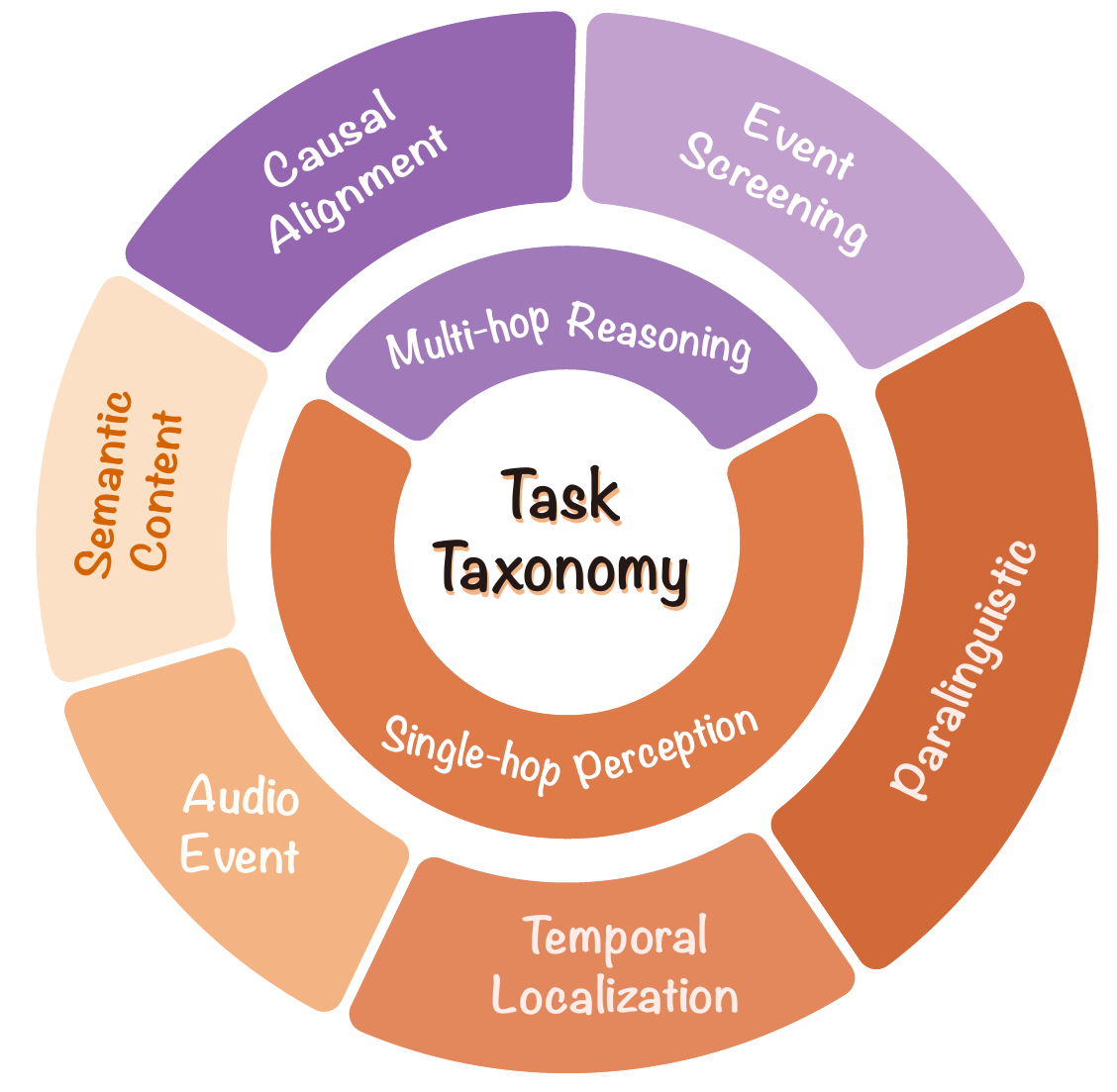}
    % \fbox{\rule[-.5cm]{0cm}{3.0cm} \rule[-.5cm]{4.2cm}{0cm}}
    \caption{Task Taxonomy}
    \label{fig:task_sunburst1}
  \end{subfigure}\hfill
  \begin{subfigure}[b]{0.32\textwidth}
    \centering
    \includegraphics[width=0.98\linewidth]{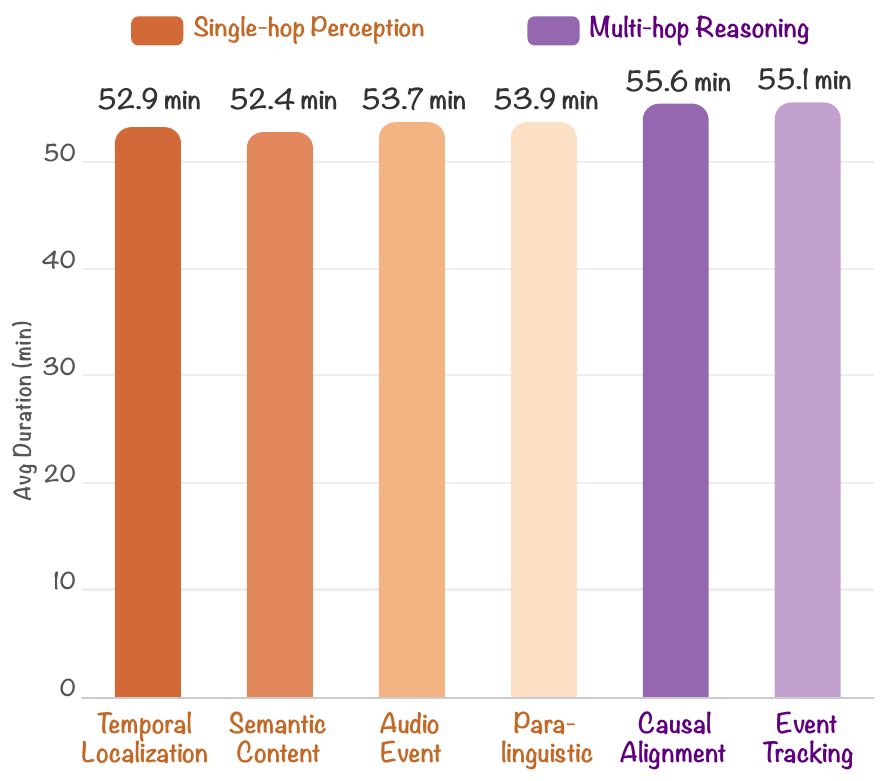}
    % \fbox{\rule[-.5cm]{0cm}{3.0cm} \rule[-.5cm]{4.2cm}{0cm}}
    \caption{Duration Statistics across Tasks}
    \label{fig:task_sunburst2}
  \end{subfigure}\hfill
  % \vspace{-0.15cm}
  \caption{(\textbf{a}) Data distribution across five domains encompassing modalities of speech, sound, and music events. (\textbf{b}) Task taxonomy partitioning evaluation into foundational single-hop perception and complex multi-hop reasoning tiers. (\textbf{c}) Statistical consistency of hour-scale durations across diverse task types to ensure balanced evaluation in duration.}
  \label{fig:combined_stat}
  \vspace{-0.cm}
\end{figure}

\benchname is a benchmark designed to evaluate the long-context understanding of LALMs. Figure~\ref{fig:combined_stat} and Table~\ref{tab:compare_prior_b} summarize the specific statistics across data and task, while Table~\ref{tab:main_results} presents evaluation results across state-of-the-art LALMs.
% AS illustrated in Table~\ref{tab:compare_prior_b}, we collect 123 long-form recordings paired with 1500 hierarchically organized multiple-choice items spanning two understanding tiers. 

% \vspace{-1.5mm}
\paragraph{Domain Coverage.}
Unlike prior benchmarks that artificially splice short clips~\cite{longspeech,audiomarathon}, \benchname~preserves entire long-form recordings spanning authentic real-world scenarios across five diverse domains in English and Chinese. As illustrated in Figure~\ref{fig:modality_pie}, our corpus encompasses: \emph{(i) Sports Commentary}, featuring rapid play-by-play narration with dense temporal events; \emph{(ii) E-sports Casting}, characterized by overlapping team communications, and chaotic multi-speaker environments; \emph{(iii) TV Dramas}, presenting complex character interactions, emotional arcs, and rich background soundscapes; \emph{(iv) News Broadcasts}, offering structured yet information-dense content with formal register and topic transitions; and \emph{(v) Interview Podcasts}, exhibiting natural conversational flow, turn-taking dynamics, and extended argumentative discourse.

\paragraph{Task Coverage.}
As illustrated in Figure~\ref{fig:teaser} and~\ref{fig:task_sunburst1}, we organize questions 
into a two-tier taxonomy: 
\textbf{Tier-1} establishes foundational audio-language understanding through four perceptual pillars: \emph{(i) Temporal Localization} requires precise timestamp retrieval of events within hour-long streams, testing fine-grained temporal indexing; \textit{(ii) Semantic Content} probes factual comprehension and topic tracking across extended discourse; \textit{(iii) Audio/Acoustic Events} evaluates non-speech sound and music recognition; and \textit{(iv) Paralinguistic Analysis} assesses speaker characteristic understanding (emotion, age, gender, timbre, pitch) beyond its transcription. 
\textbf{Tier-2} raises the challenge to multi-hop reasoning over non-contiguous segments, covering two complementary patterns. 
\emph{Event Tracking} serves as a stress test for long-term memory, requiring models to exhaustively scan the audio and aggregate sparse, often implicit events, such as counting how many times an event occurs or determining which event never happens. 
In contrast, \emph{Causal Alignment} evaluates logical coherence and narrative reasoning, requiring models to reconstruct causal chains from distributed evidence, identify temporal dependencies, and detect logical inconsistencies.

Together, these six task categories comprehensively evaluate the capabilities required for hour-scale audio understanding, ranging from low-level perceptual analysis to high-level reasoning, and from local temporal grounding to global narrative aggregation.
% \vspace{-0.2cm}
\subsection{Data Curation Pipeline}
\label{sec:pipeline_}

\begin{figure}[t]
    \centering
    \includegraphics[width=0.99\linewidth]{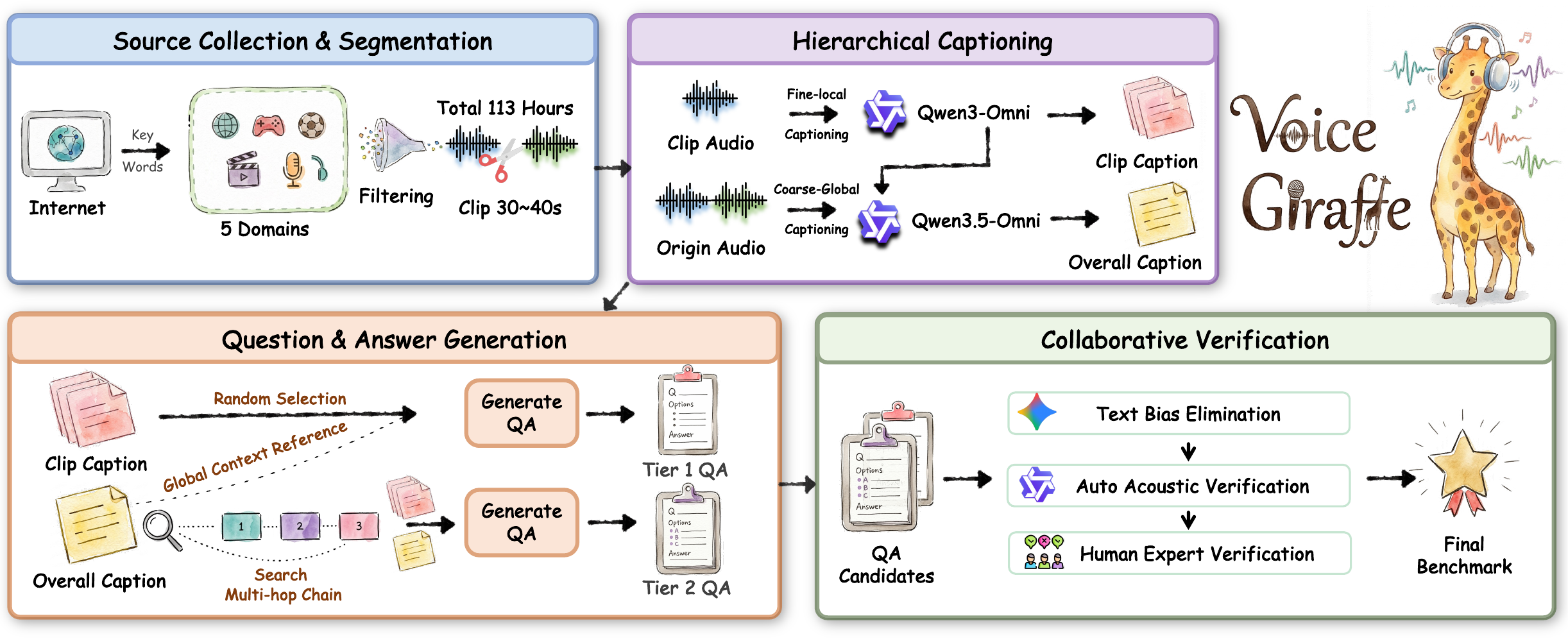}
    % \fbox{\rule[-.5cm]{0cm}{2.6cm} \rule[-.5cm]{14cm}{0cm}}
    \caption{The four-stage data construction pipeline of \benchname:
    (1)~source collection and VAD-guided segmentation;
    (2)~hierarchical captioning with state-of-the-art Omni model;
    (3)~question and answer construction across two tiers;
    (4)~collaborative model-to-human verification.}
    \label{fig:pipeline}
    % \vspace{-0.4cm}
\end{figure}

As shown in Figure~\ref{fig:pipeline}, We adopt a rigorous multi-stage framework to construct
\benchname that transforms raw long-form recordings into audited QA pairs through Voice Activity Detection (VAD) segmentation, clip-level captioning, overall captioning, QA generation, and multimodal verification. 

\noindent\textbf{Step 1. Source Collection and Segmentation.}
We collect complete long-form recordings from public platforms across five domains. For each domain, we use keyword lists to retrieve content. Retrieved recordings undergo manual filtering to exclude those outside 30 minutes to 2 hours, or with domain misalignment. Retained recordings are segmented into 30--40-second clips using \texttt{pyannote} VAD~\cite{pyannote}, establishing global time anchors for consistent annotation across all stages.

\noindent\textbf{Step 2. Hierarchical Captioning.}
Each clip is captioned by \texttt{Qwen3-Omni}~\cite{qwen3omni} covering voices (timestamp, speaker ID, transcript, paralinguistic cues), sound effects (category, timestamp), and background music. To further capture dependencies beyond individual clips, we then input all clip captions together with the original long-form audio into \texttt{Qwen3.5-Omni}~\cite{qwen3.5omni}, generating a holistic caption for building global themes, narrative structure, and cross-segment event relationships. This two-tier design enables both local detail and global context for question construction.

\noindent\textbf{Step 3. QA Construction.}
Based on the annotated metadata, we employ \texttt{Gemini-3.0}~\cite{gemini} with an evidence-grounded pipeline. 
For Tier-1 perception tasks, we pair the target clip caption with its overall caption, ensuring that fine-grained perceptual questions are grounded in local evidence while remaining consistent with the global context reference.
For Tier-2 reasoning tasks, we adopt a two-stage synthesis process. We first identify potential multi-hop reasoning pathways from the overall caption, and then retrieve the corresponding clip captions to anchor the final QA generation.
This retrieval-anchored strategy reduces hallucination by ensuring that complex reasoning chains are supported by verified clip-level evidence. 
Each question is paired with four multiple-choice options, including adversarial distractors designed to target partial reasoning shortcuts.

\noindent\textbf{Step 4. Collaborative Verification.}
Each candidate question undergoes a three-stage collaborative filtering process, in which specialized models and human experts verify complementary quality dimensions.
\emph{（i) Text-bias elimination}: a text-only LLM assesses whether the question can be answered without listening to the audio to discard audio-independent items. 
\emph{(ii) Automated acoustic verification}: an omni-modal model checks whether each QA pair is grounded in specific acoustic evidence from the corresponding audio clips, where questions without verifiable audio cues are removed.
\emph{(iii) Human expert review}: eight human annotators independently review the benchmark to ensure that each question is grounded in the audio, answerable from the provided evidence, and paired with an accurate designated answer.
Only questions that pass all three stages are included in the final benchmark.

\section{Experiments}
\label{sec:experiments}
\vspace{-2mm}
\noindent\textbf{Benchmark Models.}
% \vspace{-1mm}
% \label{sec:exp_models}
We benchmark a comprehensive suite of contemporary models against human  performance as an upper-bound reference. The suite comprises two model families: \emph{large audio language models} designed for audio-language understanding including MiMo-Audio~\cite{mimo-audio}, MOSS-Audio~\cite{mossaudio}, Audio-Flamingo-3~\cite{audio-flamingo-3}, and \emph{omni-modal language models} supporting fully multimodal input/output including QwenOmni series~\cite{qwen2omni,qwen3omni,qwen3.5omni}, Gemini series~\cite{gemini}, Phi-4-Multimodal~\cite{phi4}, MiniCPM-o-4.5~\cite{minicpm}.
Furthermore, we introduce a reasoning-enhanced cascading setting, where the Large Reasoning Model (LRM) \texttt{GPT-5.2-Thinking}~\cite{openai2025gpt52systemcard} performs inference-time reasoning over clip-level audio captions generated by above LALMs to answer the corresponding QA pairs. 

\begin{table*}[t]
\centering
\caption{\benchname results across 14 models and 6 task categories: 4 single-hop tasks---\textit{Temp.~Loc.}~(Temporal Localization), \textit{Sem.~Con.}~(Semantic Content), \textit{Aco.~Evt.}~(Acoustic Event), and \textit{Paralin.}~(Paralinguistic), and 2 multi-hop tasks---\textit{Causal} (Causal alignment) and \textit{Evt.~Trk.}~(Event Tracking). \textbf{Size}: number of activated parameters. \textbf{{E2E}}: end-to-end audio inference without caption cascading. \textbf{Think}: native extended thinking mode. \textbf{LRM}: augmentation for inference with an external large reasoning model over streaming clip audio captions. Best results are highlighted in setting-specific \textbf{bold} colors: \bestE{\textbf{red}} for end-to-end inference, \bestC{\textbf{yellow}} for cascaded caption aggregation, and \bestR{\textbf{blue}} for reasoning-enhanced cascading.}
\label{tab:main_results}
\resizebox{\linewidth}{!}{
\begin{tabular}{l c cc c cccc c cc c c}
    \toprule
    \toprule
    \multirow{2.5}{*}{\textbf{Model}} & \multirow{2.5}{*}{\textbf{Size}} & \multirow{2.5}{*}{\textbf{Thinking}} & \multirow{2.5}{*}{\textbf{E2E}} & \multirow{2.5}{*}{\textbf{LRM}} & \multicolumn{5}{c}{\textbf{Single-hop Perception}} & \multicolumn{3}{c}{\textbf{Multi-hop Reasoning}} & \multirow{1.6}{*}{\textbf{Overall}} \\
    \cmidrule(lr){6-10} \cmidrule(lr){11-13}
    & & & & & Temp. Loc. & Sem. Con. & Aco. Evt. & Paralin. & Avg. & Causal & Evt. Trk. & Avg. & \multirow{0.9}{*}{\textbf{Avg.}} \\
    \midrule
    \midrule
    \multicolumn{14}{c}{\textbf{Open-source Models}} \\
    \midrule
    \midrule
    
    \multirow{2}{*}{MiniCPM-o-4.5} & \multirow{2}{*}{9B} & \multirow{2}{*}{\ding{55}} & \multirow{2}{*}{\ding{55}} & \ding{55}  & \cellcolor{casbg}3.60 & \cellcolor{casbg}4.00 & \cellcolor{casbg}4.50 & \cellcolor{casbg}2.29 & \cellcolor{casbg}3.40 & \cellcolor{casbg}5.20 & \cellcolor{casbg}3.60 & \cellcolor{casbg}4.40 & \cellcolor{casbg}3.73 \\
    & & & & \ding{51} & \cellcolor{reasbg}69.60 & \cellcolor{reasbg}91.00 & \cellcolor{reasbg}51.00 & \cellcolor{reasbg}68.86 & \cellcolor{reasbg}69.90 & \cellcolor{reasbg}42.40 & \cellcolor{reasbg}55.60 & \cellcolor{reasbg}49.00 & \cellcolor{reasbg}\bestR{62.93} \\
    \cdashline{1-14}
    \noalign{\vskip 1mm}
    \multirow{2}{*}{Phi-4-Multimodal} & \multirow{2}{*}{5.6B} & \multirow{2}{*}{\ding{55}} & \multirow{2}{*}{\ding{55}} & \ding{55} & \cellcolor{casbg}30.40 & \cellcolor{casbg}30.00 & \cellcolor{casbg}31.50 & \cellcolor{casbg}33.71 & \cellcolor{casbg}31.70 & \cellcolor{casbg}30.00 & \cellcolor{casbg}23.60 & \cellcolor{casbg}26.80 & \cellcolor{casbg}30.07 \\
    & & & & \ding{51}  & \cellcolor{reasbg}32.80 & \cellcolor{reasbg}39.50 & \cellcolor{reasbg}44.00 & \cellcolor{reasbg}50.86 & \cellcolor{reasbg}42.70 & \cellcolor{reasbg}39.60 & \cellcolor{reasbg}22.80 & \cellcolor{reasbg}31.20 & \cellcolor{reasbg}38.87 \\
    \cdashline{1-14}
    \noalign{\vskip 1mm}
    \multirow{2}{*}{Audio-Flamingo-3} & \multirow{2}{*}{8B} & \multirow{2}{*}{\ding{55}} & \multirow{2}{*}{\ding{55}} & \ding{55} & \cellcolor{casbg}35.60 & \cellcolor{casbg}43.50 & \cellcolor{casbg}40.50 & \cellcolor{casbg}42.29 & \cellcolor{casbg}40.50 & \cellcolor{casbg}64.40 & \cellcolor{casbg}40.40 & \cellcolor{casbg}52.40 & \cellcolor{casbg}44.47 \\
    & & & & \ding{51}  & \cellcolor{reasbg}65.60 & \cellcolor{reasbg}52.00 & \cellcolor{reasbg}49.00 & \cellcolor{reasbg}58.86 & \cellcolor{reasbg}57.20 & \cellcolor{reasbg}38.40 & \cellcolor{reasbg}33.60 & \cellcolor{reasbg}36.00 & \cellcolor{reasbg}50.13 \\
    \cdashline{1-14}
    \noalign{\vskip 1mm}
    \multirow{2}{*}{Qwen2.5-Omni} & \multirow{2}{*}{7B} & \multirow{2}{*}{\ding{55}} & \multirow{2}{*}{\ding{55}} & \ding{55} & \cellcolor{casbg}38.00 & \cellcolor{casbg}48.00 & \cellcolor{casbg}31.50 & \cellcolor{casbg}50.57 & \cellcolor{casbg}43.10 & \cellcolor{casbg}43.20 & \cellcolor{casbg}30.80 & \cellcolor{casbg}37.00 & \cellcolor{casbg}41.07 \\
    & & & & \ding{51} & \cellcolor{reasbg}74.40 & \cellcolor{reasbg}92.00 & \cellcolor{reasbg}39.50 & \cellcolor{reasbg}62.86 & \cellcolor{reasbg}66.90 & \cellcolor{reasbg}41.60 & \cellcolor{reasbg}52.40 & \cellcolor{reasbg}47.00 & \cellcolor{reasbg}60.27 \\
    \cdashline{1-14}
    \noalign{\vskip 1mm}
    \multirow{2}{*}{MiMo-Audio} & \multirow{2}{*}{7B} & \multirow{2}{*}{\ding{55}} & \multirow{2}{*}{\ding{55}} & \ding{55} & \cellcolor{casbg}34.00 & \cellcolor{casbg}39.50 & \cellcolor{casbg}43.50 & \cellcolor{casbg}46.29 & \cellcolor{casbg}41.30 & \cellcolor{casbg}40.40 & \cellcolor{casbg}23.60 & \cellcolor{casbg}32.00 & \cellcolor{casbg}38.20 \\
    & & & & \ding{51} & \cellcolor{reasbg}\bestR{84.00} & \cellcolor{reasbg}88.00 & \cellcolor{reasbg}48.50 & \cellcolor{reasbg}65.43 & \cellcolor{reasbg}\bestR{71.20} & \cellcolor{reasbg}42.40 & \cellcolor{reasbg}50.00 & \cellcolor{reasbg}46.20 & \cellcolor{reasbg}{62.87} \\
    \cdashline{1-14}
    \noalign{\vskip 1mm}
    \multirow{4}{*}{MOSS-Audio} & \multirow{4}{*}{8B} & \multirow{2}{*}{\ding{55}} & \multirow{2}{*}{\ding{55}} & \ding{55} & \cellcolor{casbg}28.80 & \cellcolor{casbg}40.50 & \cellcolor{casbg}39.00 & \cellcolor{casbg}47.43 & \cellcolor{casbg}39.70 & \cellcolor{casbg}35.20 & \cellcolor{casbg}26.00 & \cellcolor{casbg}30.60 & \cellcolor{casbg}36.67 \\
    & & & & \ding{51} & \cellcolor{reasbg}62.40 & \cellcolor{reasbg}72.50 & \cellcolor{reasbg}51.00 & \cellcolor{reasbg}64.86 & \cellcolor{reasbg}63.00 & \cellcolor{reasbg}42.00 & \cellcolor{reasbg}40.40 & \cellcolor{reasbg}41.20 & \cellcolor{reasbg}55.73 \\
    \cdashline{3-14}
    \noalign{\vskip 0.8mm}
    &  & \multirow{2}{*}{\ding{51}} & \multirow{2}{*}{\ding{55}} & \ding{55} & \cellcolor{casbg}31.20 & \cellcolor{casbg}36.00 & \cellcolor{casbg}44.00 & \cellcolor{casbg}60.00 & \cellcolor{casbg}44.80 & \cellcolor{casbg}65.60 & \cellcolor{casbg}27.20 & \cellcolor{casbg}46.40 & \cellcolor{casbg}45.33 \\
    & & & & \ding{51} & \cellcolor{reasbg}73.20 & \cellcolor{reasbg}76.50 & \cellcolor{reasbg}52.50 & \cellcolor{reasbg}67.43 & \cellcolor{reasbg}67.70 & \cellcolor{reasbg}40.80 & \cellcolor{reasbg}41.60 & \cellcolor{reasbg}41.20 & \cellcolor{reasbg}58.87 \\
    \cdashline{1-14}
    \noalign{\vskip 1mm}

    \multirow{4}{*}{Qwen3-Omni} & \multirow{4}{*}{30B(A3B)} & \multirow{2}{*}{\ding{55}} & \multirow{2}{*}{\ding{55}} & \ding{55} & \cellcolor{casbg}30.80 & \cellcolor{casbg}69.00 & \cellcolor{casbg}48.50 & \cellcolor{casbg}61.43 & \cellcolor{casbg}52.70 & \cellcolor{casbg}45.60 & \cellcolor{casbg}47.20 & \cellcolor{casbg}46.40 & \cellcolor{casbg}50.60 \\
    & & &  & \ding{51}  & \cellcolor{reasbg}39.20 & \cellcolor{reasbg}91.50 & \cellcolor{reasbg}51.00 & \cellcolor{reasbg}68.29 & \cellcolor{reasbg}62.20 & \cellcolor{reasbg}46.00 & \cellcolor{reasbg}52.40 & \cellcolor{reasbg}49.20 & \cellcolor{reasbg}57.87 \\
    \cdashline{3-14}
    \noalign{\vskip 0.8mm}
    &  & \multirow{2}{*}{\ding{51}} &  \multirow{2}{*}{\ding{55}} & \ding{55}  & \cellcolor{casbg}20.00 & \cellcolor{casbg}58.00 & \cellcolor{casbg}43.00 & \cellcolor{casbg}46.57 & \cellcolor{casbg}41.50 & \cellcolor{casbg}63.20 & \cellcolor{casbg}36.00 & \cellcolor{casbg}49.60 & \cellcolor{casbg}44.20 \\
    & & &  & \ding{51} & \cellcolor{reasbg}38.80 & \cellcolor{reasbg}68.50 & \cellcolor{reasbg}50.50 & \cellcolor{reasbg}64.29 & \cellcolor{reasbg}56.00 & \cellcolor{reasbg}39.20 & \cellcolor{reasbg}42.40 & \cellcolor{reasbg}40.80 & \cellcolor{reasbg}50.93 \\
    \midrule
    \midrule
    \multicolumn{14}{c}{\textbf{Proprietary Models}} \\
    \midrule
    \midrule
    \multirow{3}{*}{Gemini-2.5-Pro} & \multirow{3}{*}{--} & \multirow{3}{*}{\ding{55}} & \ding{55} & \ding{55} & \cellcolor{casbg}72.80 & \cellcolor{casbg}95.50 & \cellcolor{casbg}58.00 & \cellcolor{casbg}68.29 & \cellcolor{casbg}72.80 & \cellcolor{casbg}74.40 & \cellcolor{casbg}\bestC{70.00} & \cellcolor{casbg}72.20 & \cellcolor{casbg}72.60 \\
     & & & \ding{55}   & \ding{51} & \cellcolor{reasbg}37.60 & \cellcolor{reasbg}95.50 & \cellcolor{reasbg}59.50 & \cellcolor{reasbg}\bestR{70.57} & \cellcolor{reasbg}65.10 & \cellcolor{reasbg}\bestR{46.80} & \cellcolor{reasbg}60.40 & \cellcolor{reasbg}\bestR{53.60} & \cellcolor{reasbg}61.27 \\
    & & & \ding{51} & \ding{55}  & \cellcolor{e2ebg}33.20 & \cellcolor{e2ebg}61.00 & \cellcolor{e2ebg}43.50 & \cellcolor{e2ebg}52.57 & \cellcolor{e2ebg}47.60 & \cellcolor{e2ebg}52.00 & \cellcolor{e2ebg}37.20 & \cellcolor{e2ebg}44.60 & \cellcolor{e2ebg}46.60 \\
    \cdashline{1-14}
    \noalign{\vskip 1mm}
    \multirow{3}{*}{Gemini-3.1-Pro} & \multirow{3}{*}{--} & \multirow{3}{*}{\ding{55}} & \ding{55} & \ding{55} & \cellcolor{casbg}\bestC{79.60} & \cellcolor{casbg}\bestC{97.50} & \cellcolor{casbg}{57.00} & \cellcolor{casbg}66.29 & \cellcolor{casbg}\bestC{74.00} & \cellcolor{casbg}\bestC{85.20} & \cellcolor{casbg}69.60 & \cellcolor{casbg}\bestC{77.40} & \cellcolor{casbg}\bestC{75.13} \\
     & & & \ding{55}   & \ding{51} & \cellcolor{reasbg}32.80 & \cellcolor{reasbg}\bestR{96.50} & \cellcolor{reasbg}57.00 & \cellcolor{reasbg}68.00 & \cellcolor{reasbg}62.70 & \cellcolor{reasbg}45.60 & \cellcolor{reasbg}\bestR{61.20} & \cellcolor{reasbg}53.40 & \cellcolor{reasbg}59.60 \\
    & & & \ding{51} & \ding{55}  & \cellcolor{e2ebg}32.80 & \cellcolor{e2ebg}53.00 & \cellcolor{e2ebg}33.50 & \cellcolor{e2ebg}46.29 & \cellcolor{e2ebg}41.70 & \cellcolor{e2ebg}36.40 & \cellcolor{e2ebg}40.00 & \cellcolor{e2ebg}38.20 & \cellcolor{e2ebg}40.53 \\
    \cdashline{1-14}
    \noalign{\vskip 1mm}
    \multirow{3}{*}{Qwen3.5-Omni-Flash} & \multirow{3}{*}{--} & \multirow{3}{*}{\ding{55}} & \ding{55} & \ding{55}  & \cellcolor{casbg}31.60 & \cellcolor{casbg}79.50 & \cellcolor{casbg}60.00 & \cellcolor{casbg}64.57 & \cellcolor{casbg}58.40 & \cellcolor{casbg}64.00 & \cellcolor{casbg}49.60 & \cellcolor{casbg}56.80 & \cellcolor{casbg}57.87 \\
    & & & \ding{55}   & \ding{51} & \cellcolor{reasbg}37.20 & \cellcolor{reasbg}83.00 & \cellcolor{reasbg}57.00 & \cellcolor{reasbg}68.29 & \cellcolor{reasbg}61.20 & \cellcolor{reasbg}45.20 & \cellcolor{reasbg}52.00 & \cellcolor{reasbg}48.60 & \cellcolor{reasbg}57.00 \\
    & & & \ding{51} & \ding{55}  & \cellcolor{e2ebg}72.80 & \cellcolor{e2ebg}84.50 & \cellcolor{e2ebg}56.50 & \cellcolor{e2ebg}59.71 & \cellcolor{e2ebg}67.30 & \cellcolor{e2ebg}70.00 & \cellcolor{e2ebg}47.60 & \cellcolor{e2ebg}58.80 & \cellcolor{e2ebg}64.47 \\
    \cdashline{1-14}
    \noalign{\vskip 1mm}
    \multirow{3}{*}{Qwen3.5-Omni-Plus} & \multirow{3}{*}{--} & \multirow{3}{*}{\ding{55}} & \ding{55} & \ding{55} & \cellcolor{casbg}44.80 & \cellcolor{casbg}92.00 & \cellcolor{casbg}\bestC{60.50} & \cellcolor{casbg}\bestC{69.43} & \cellcolor{casbg}66.00 & \cellcolor{casbg}76.40 & \cellcolor{casbg}56.80 & \cellcolor{casbg}66.60 & \cellcolor{casbg}66.20 \\
     & & & \ding{55}   & \ding{51} & \cellcolor{reasbg}39.60 & \cellcolor{reasbg}92.50 & \cellcolor{reasbg}\bestR{64.00} & \cellcolor{reasbg}67.43 & \cellcolor{reasbg}64.80 & \cellcolor{reasbg}42.80 & \cellcolor{reasbg}55.20 & \cellcolor{reasbg}49.00 & \cellcolor{reasbg}59.53 \\
    & & & \ding{51} & \ding{55}  & \cellcolor{e2ebg}\bestE{92.00} & \cellcolor{e2ebg}\bestE{94.00} & \cellcolor{e2ebg}\bestE{64.00} & \cellcolor{e2ebg}\bestE{72.00} & \cellcolor{e2ebg}\bestE{79.80} & \cellcolor{e2ebg}\bestE{77.20} & \cellcolor{e2ebg}\bestE{59.60} & \cellcolor{e2ebg}\bestE{68.40} & \cellcolor{e2ebg}\bestE{76.00} \\
   
    \midrule
    \multicolumn{5}{l}{\textit{Human Reference}}  & 63.89 & 90.74 & 79.63 & 79.17 & 77.38 & 52.24 & 61.11 & 56.20 & 70.51 \\
    \bottomrule
    \bottomrule
\end{tabular}
\vspace{-2mm}
}
\vspace{-2mm}
\end{table*}

% \bestE
% \bestC
% \bestR

\noindent\textbf{Inference Settings.}
We implement three inference settings to evaluate models with different context capacities.
\emph{(i) End-to-End (E2E).}
For models that natively support hour-scale context windows, we directly feed the complete audio signal together with the query for inference.
\emph{(ii) Cascaded Caption Aggregation.}
For models limited to short audio inputs, we adopt a cascaded pipeline that performs sliding-window captioning over the full audio timeline with a 30-second window size. Each window is captioned with speech transcripts, speaker information (gender, age, emotion, and pitch etc.), sound events, and music cues. The resulting clip-level captions are concatenated into a textual description and processed together with the query by original LALM.
\emph{(iii) Reasoning-Enhanced Cascading.}
This setting extends the cascaded pipeline by feeding the aggregated captions of LALMs and the query into the stronger large reasoning model~\cite{openai2025gpt52systemcard}, enabling inference-time reasoning over the audio caption.

\noindent\textbf{Evaluation Settings.}
Given that all \benchname questions follow a multiple-choice format, we adopt macro-average accuracy as the primary metric. Specifically, each model receives the audio recording, question text, and candidate options as input, and we assess whether its selected option matches the ground truth. Following established practices in MMAU~\cite{mmau} and MMAR~\cite{mmar}, we employ regular expression matching and string alignment to extract and compare model predictions against gold labels. For models without explicit reasoning output, we directly evaluate the final output. For models with a thinking mode that generate an intermediate chain-of-thought steps, we strip the reasoning content and evaluate only the final answer, ensuring fair and consistent comparison across different model architectures.

\noindent\textbf{Human Reference.} We engage eight annotators with native-level proficiency in both English and Chinese. Given the hour-scale listening burden, we improve efficiency by sampling multiple questions per audio recording across all six tasks, allowing comprehensive evaluation from a single listening. Sampling is stratified across domains and languages. Each annotator independently evaluates a randomly assigned subset of 150 questions (75 per language), sampled via stratified random sampling to preserve the benchmark's proportional distribution across domains and languages.

\subsection{Experimental Results}
\label{sec:exp_main}

Table~\ref{tab:main_results} presents model performance of \benchname, evaluated via multiple-choice accuracy (\%). Results reveal several critical findings regarding the benchmark's difficulty and LALM's capabilities.

% \begin{wrapfigure}{r}{0.44\textwidth}
%   \centering
%   \vspace{-0.3cm}
%   \includegraphics[width=0.99\linewidth]{Figures/fig_sem_aco_delta.pdf}
%   \vspace{-0.3cm}
%   \caption{Semantic-task declines while acoustic-task rises with longer audio.}
%   \label{fig:fig_sem_aco_delta}
%   \vspace{-0.4cm}
% \end{wrapfigure}
\noindent\textbf{Finding 1: Hour-scale understanding of \benchname remains extremely challenging for all models.} Among the four models capable of hour-scale E2E inference, only Qwen3.5-Omni-Plus (76.00\%) surpasses the human reference of 70.51\%, while the remaining three fall substantially below human performance.
For open-source models that rely solely on cascaded caption aggregation without LRM enhancement, the best overall accuracy reaches merely 50.60\% (Qwen3-Omni), with most models clustered between 30\% and 45\%. 
However, after applying reasoning-enhanced cascading, overall performance improves markedly across models, suggesting that current LALMs possess reasonable short-term perception ability but still face major bottlenecks in long-term reasoning and memory aggregation.
Notably, even for human annotators, it presents substantial difficulty, requiring both sparse event localization within long-context and sustained granular memory retention across extended temporal spans. This wide performance gap underscores that hour-scale audio understanding remains a formidable open problem and confirms \benchname~as a challenging benchmark that is far from saturation.

\begin{wrapfigure}{r}{0.3\textwidth}
  \centering
  \vspace{-0.5cm}
  \includegraphics[width=1.0\linewidth]{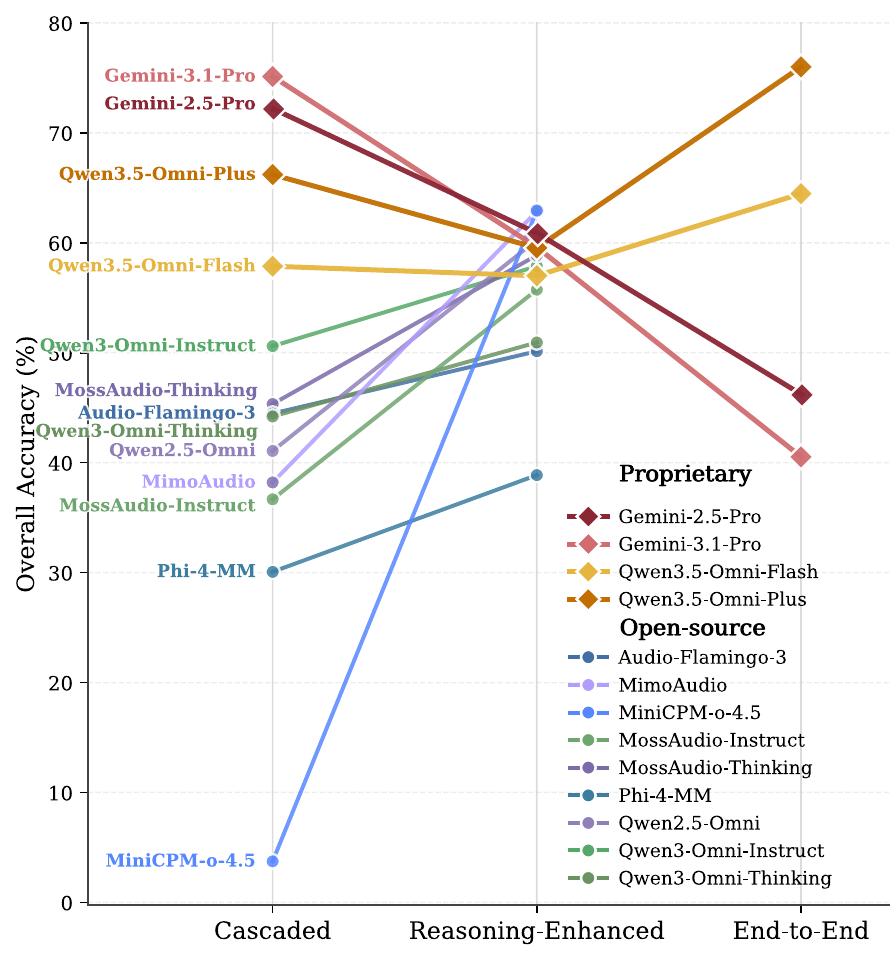}
  % \vspace{-0.3cm}
  \caption{Comparison of performance across inference modes.}
  \label{fig:fig_fingding2}
  \vspace{-0.67cm}
\end{wrapfigure}
\noindent\textbf{Finding 2: The optimal inference paradigm is model-dependent.}
As shown in Figure~\ref{fig:fig_fingding2}, the three inference settings show distinct strengths across model families. 
For models with strong native long-context audio understanding, E2E inference is most effective. Qwen3.5-Omni-Plus achieves the best overall score of 76.00\%, outperforming its cascaded variant by 9.8\% while Qwen3.5-Omni-Flash also gains 6.6\% from E2E inference.
This suggests that direct inference can better preserve fine-grained perceptual cues when the LALMs is sufficiently robust for the long-context.
However, this pattern does not hold universally. Gemini-3.1-Pro performs much better with cascaded caption aggregation than with E2E inference, dropping 34.6\% under direct audio input, indicating that hour-scale context can overwhelm certain audio understanding. 

Reasoning-enhanced cascading exhibits a complementary but asymmetric effect.
For open-source models, external LRM reasoning substantially improves performance, raising the average score from 37.15\% to 55.39\%.
This suggests that open-source LALMs often generate captions with usable evidence but lack sufficient long-range text aggregation ability.
In contrast, for proprietary models with stronger long-context reasoning capacities, LRM augmentation can substantially degrade performance, indicating that the external LRM itself may become a bottleneck when its reasoning ability falls behind that of the evaluated closed-source model. This highlights the importance of LRM selection and motivates our further comparison of more advanced MLLMs as reasoning backbones in Sec.~\ref{subsec:LRM_selection}.

\begin{wrapfigure}{r}{0.3\textwidth}
  \centering
  \vspace{-0.5cm}
  \includegraphics[width=1.0\linewidth]{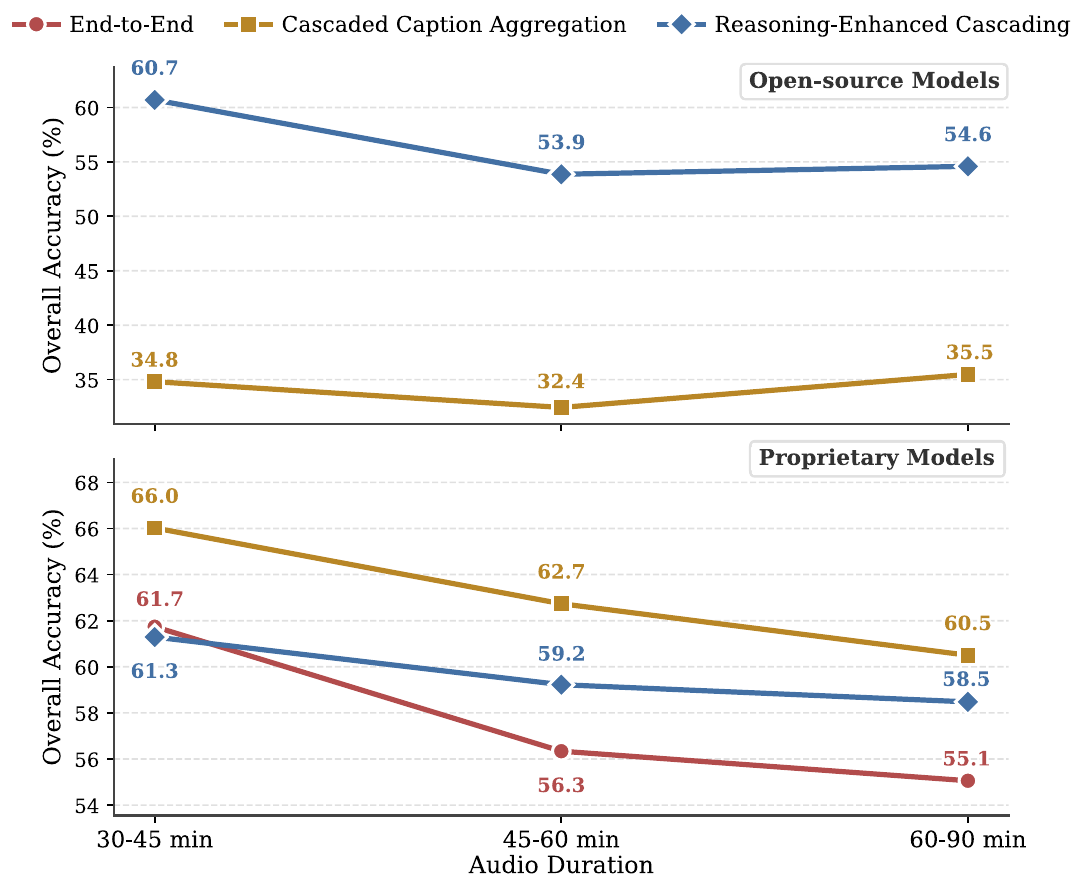}
  % \vspace{-0.3cm}
  \caption{Comparison of performance across duration.}
  \label{fig:fig_duration}
  \vspace{-0.67cm}
\end{wrapfigure}
\noindent\textbf{Finding 3: Increasing audio duration systematically amplifies understanding bottlenecks.} As shown in Figure~\ref{fig:fig_duration}, extending the audio context universally degrades performance across all model families, yet the failure mechanisms differ significantly by inference mode. For open-source models, native Cascaded Caption Aggregation effectively bottoms out at 32\%--36\%, lacking the capacity to handle long-context understanding. Introducing an external LRM effectively resuscitates performance but still exhibits a clear decay from 60.7\% to roughly 54\%. Conversely, for proprietary models, Cascaded Caption Aggregation remains the strongest mode in absolute accuracy, but still declines markedly from 66.0\% to 60.5\% as duration increases.
E2E inference exhibits a comparable degradation, dropping from 61.7\% to 55.1\%, suggesting that both ways are increasingly challenged by longer inputs.
By contrast, Reasoning-Enhanced Cascading shows the smallest drop, decreasing only from 61.3\% to 58.5\%, indicating that the external LRM offers more stable long-context reasoning than LALMs themselves, although it does not always yield the highest absolute accuracy.
Overall, the downward trajectories across all three modes confirm that current models still struggle increasingly as audio duration grows.

% \begin{wrapfigure}{l}{0.34\textwidth}
%   \centering
%   \vspace{-0.5cm}
%   \includegraphics[width=1.0\linewidth]{Figures/radar_t2.pdf}
%   \vspace{-0.3cm}
%   \caption{Semantic-task declines while acoustic-task rises with longer audio.}
%   \label{fig:fig_sem_aco_delta}
%   \vspace{-0.4cm}
% \end{wrapfigure}

\begin{figure}[t]
    \centering
    \includegraphics[width=0.90\linewidth]{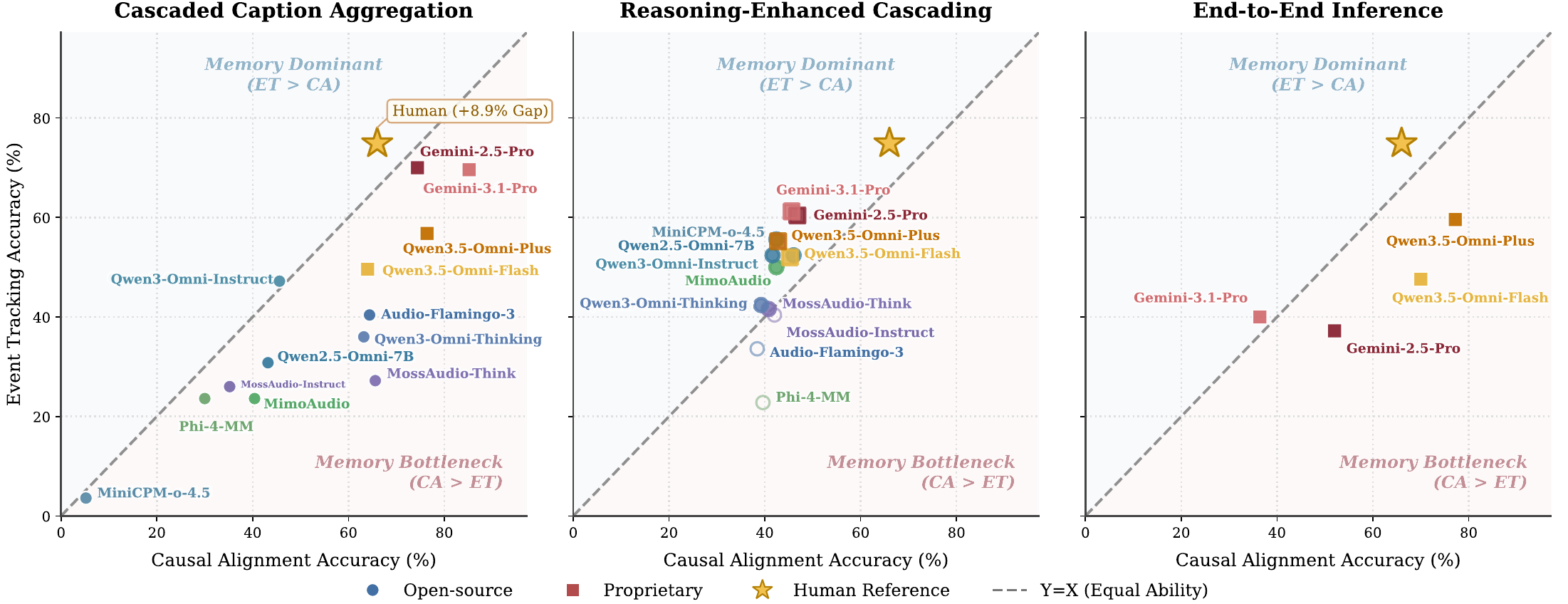}
    % \vspace{-3.5mm}
    \caption{Performance asymmetry between Causal Alignment (CA) and Event Tracking (ET). The diagonal ($Y=X$) represents balanced capability. Current LALMs across both native Cascaded and E2E settings overwhelmingly fall into the lower "Memory Bottleneck" region ($CA > ET$). }
    \label{fig:finding4}
    % \vspace{-0.52cm}
\end{figure}

\textbf{Finding 4: Long-range memory remains a key bottleneck, in contrast to human.} 
As shown in Figure~\ref{fig:finding4}, models across both cascaded and E2E settings consistently achieve higher scores on Causal Alignment than on Event Tracking, with gaps of 4--20\% for proprietary models and up to 38\% for open-source models. This trend contrasts sharply with human performance, where annotators score 8.9\% higher on Event Tracking than on Causal Alignment, reflecting stronger ability of human to maintain persistent episodic memory over long-term spans. 
The asymmetry suggests that current LALMs are relatively capable of reconstructing causal relations from salient semantic cues, but struggle to preserve and aggregate sparse event states across hour-scale contexts. 
In other words, their failures are less about causal reasoning and more about persistent memory over audio evidence across long-term contexts. 
Notably, LRM enhancement partially mitigates this imbalance.
Under reasoning-enhanced cascading, Event Tracking often surpasses Causal Alignment, suggesting that strong LRMs possess more robust long-context memory and evidence aggregation capabilities than LALMs themselves when reasoning over temporally ordered captions.
However, this compensation remains indirect, highlighting the need for LALMs with mechanisms for long-term memory persistent.

Overall, these findings underline the urgent need for further innovation in long-form audio understanding, to address fundamental challenges in memory persistence and acoustic perception that current models exhibit at the hour scale.

\section{Discussion}
\label{sec:analysis}

\subsection{End-to-End vs. Cascaded Inference for Hour-Level Audio Understanding}

Table~\ref{tab:main_results} reveals a clear trade-off between end-to-end and cascaded inference for hour-level audio understanding. End-to-end inference offers the most direct access to the original acoustic signal, preserving fine-grained cues such as speaker traits, prosody, background events, and temporal transitions that may be lost during captioning. This makes it particularly valuable for tasks that rely on detailed acoustic evidence or precise audio-visual grounding. However, its effectiveness strongly depends on the model's native long-context audio capacity. For models without robust hour-scale audio modeling, directly processing long audio often leads to severe information dilution, unstable temporal localization, and weak multi-hop reasoning. 
In contrast, cascaded inference provides a more scalable and structured solution for long-form audio. By converting hour-level audio into temporally ordered clip-level captions, the model can reason over a compact textual representation rather than the full acoustic sequence. This setting is especially effective for semantic content understanding, event tracking, and causal reasoning, where the key challenge is not low-level perception but locating and integrating evidence across a long timeline. The gains are more pronounced for models with limited audio context windows, suggesting that temporal caption aggregation can serve as an effective memory interface for long audio understanding. 
Nevertheless, cascaded inference introduces an information bottleneck. Once acoustic details are omitted, distorted, or hallucinated in the intermediate captions, downstream reasoning models cannot recover them. This limitation is particularly harmful for paralinguistic perception, subtle sound-event discrimination, and tasks requiring exact timing or fine-grained acoustic comparison. Reasoning-enhanced cascading further improves multi-hop and temporally dispersed questions by leveraging stronger language reasoning over the caption timeline, but it remains bounded by the quality and granularity of the generated captions.

Overall, E2E inference is preferable when models can natively handle long audio and when the task requires faithful acoustic perception. Cascaded inference is more reliable when long-context capacity is limited or when the task primarily involves semantic aggregation over extended durations. These results suggest that fine-grained perception favours direct acoustic modeling, while long-range reasoning benefits from structured captions.

\subsection{Large Reasoning Model for Enhanced Cascading Inference}
\label{subsec:LRM_selection}
% Ablation study: LRM selection under cascade caption aggregation mode.
% Helper commands for gain/loss indicators (define in preamble if not already):
% \newcommand{\up}[1]{{\scriptsize\textcolor{teal}{\textuparrow#1}}}
% \newcommand{\dn}[1]{{\scriptsize\textcolor{red}{\textdownarrow#1}}}
\begin{table*}[t]
\centering
\newcommand{\up}[1]{{\scriptsize\textcolor{teal}{$\uparrow$#1}}}
\newcommand{\dn}[1]{{\scriptsize\textcolor{red}{$\downarrow$#1}}}
\newcommand{\eqa}[1]{{\scriptsize\textcolor{blue}{#1}}}
\caption{Ablation study on LRM selection under cascade caption aggregation mode. Each LALM is paired with different large reasoning models. \textbf{No LRM} denotes the pure cascade baseline. Gains over baseline are shown as \up{} (improvement) or \dn{} (degradation).}
\label{tab:lrm_ablation}
\resizebox{\linewidth}{!}{%
\begin{tabular}{l ccc ccc ccc}
\toprule
\multirow{2}{*}{\textbf{LALM}}
 & \multicolumn{3}{c}{\textbf{No LRM} (baseline)}
 & \multicolumn{3}{c}{\textbf{GPT-5.2}}
 & \multicolumn{3}{c}{\textbf{Gemini-3.1-Pro}} \\
\cmidrule(lr){2-4} \cmidrule(lr){5-7} \cmidrule(lr){8-10}
 & Single-hop & Multi-hop & Overall
 & Single-hop & Multi-hop & Overall
 & Single-hop & Multi-hop & Overall \\
\midrule
\multicolumn{10}{l}{\textit{Open-source LALMs}} \\
\midrule
MiniCPM-o-4.5 (9B)
 & 3.40  & 4.40  & 3.70
 & 69.9 \up{66.5} & 49.0 \up{44.6} & 62.9 \up{59.2}
 & 75.8 \up{72.4} & 78.0 \up{73.6} & 76.5 \up{72.8} \\
Phi-4-Multimodal (5.6B)
 & 31.7 & 26.8 & 30.1
 & 42.7 \up{11.0} & 31.2 \up{4.4} & 38.9 \up{8.8}
 & 41.6 \up{9.9} & 29.8 \up{3.0} & 37.7 \up{7.6} \\
Audio-Flamingo-3 (8B)
 & 40.5 & 52.4 & 44.5
 & 57.2 \up{16.7} & 36.0 \dn{16.4} & 50.1 \up{5.6}
 & 57.5 \up{17.0} & 39.2 \dn{13.2} & 51.4 \up{6.9} \\
Qwen2.5-Omni (7B)
 & 43.1 & 37.0 & 41.1
 & 66.9 \up{23.8} & 47.0 \up{10.0} & 60.3 \up{19.2}
 & 71.1 \up{28.0} & 73.6 \up{36.6} & 71.9 \up{30.8} \\
MiMo-Audio (7B)
 & 41.3 & 32.0 & 38.2
 & 71.2 \up{29.9} & 46.2 \up{14.2} & 62.9 \up{24.7}
 & 74.9 \up{33.6} & 74.4 \up{42.4} & 74.7 \up{36.5} \\
MOSS-Audio (8B)
 & 39.7 & 30.6 & 36.7
 & 63.0 \up{23.3} & 41.2 \up{10.6} & 55.7 \up{19.0}
 & 68.3 \up{28.6} & 61.0 \up{30.4} & 65.9 \up{29.2} \\
MOSS-Audio-Think (8B)
 & 44.8 & 46.4 & 45.3
 & 67.7 \up{22.9} & 41.2 \dn{5.2} & 58.9 \up{13.6}
 & 74.9 \up{30.1} & 65.4 \up{19.0} & 71.7 \up{26.4} \\
Qwen3-Omni (30B/A3B)
 & 52.7 & 46.4 & 50.6
 & 62.2 \up{9.5} & 49.2 \up{2.8} & 57.9 \up{7.3}
 & 77.2 \up{24.5} & 74.8 \up{28.4} & 76.4 \up{25.8} \\
Qwen3-Omni-Think (30B/A3B)
 & 41.5 & 49.6 & 44.2
 & 56.0 \up{14.5} & 40.8 \dn{8.8} & 50.9 \up{6.7}
 & 67.2 \up{25.7} & 58.2 \up{8.6} & 64.2 \up{20.0} \\
\midrule
\multicolumn{10}{l}{\textit{Proprietary LALMs}} \\
\midrule
Gemini-2.5-Pro
 & 72.8 & 72.2 & 72.6
 & 65.1 \dn{7.7} & 53.6 \dn{18.6} & 61.3 \dn{11.3}
 & 75.4 \up{2.6} & 77.8 \up{5.6} & 76.2 \up{3.6} \\
Gemini-3.1-Pro
 & 74.0 & 77.4 & 75.1
 & 62.7 \dn{11.3} & 53.4 \dn{24.0} & 59.6 \dn{15.5}
 & 74.0 \eqa{0.0} & 77.4 \eqa{0} & 75.1 \eqa{0.0} \\
Qwen3.5-Omni-Flash
 & 58.4 & 56.8 & 57.9
 & 61.2 \up{2.8} & 48.6 \dn{8.2} & 57.0 \dn{0.9}
 & 74.6 \up{16.2} & 69.8 \up{13.0} & 73.0 \up{15.1} \\
Qwen3.5-Omni-Plus
 & 66.0 & 66.6 & 66.2
 & 64.8 \dn{1.2} & 49.0 \dn{17.6} & 59.5 \dn{6.7}
 & 79.3 \up{13.3} & 76.4 \up{9.8} & 78.3 \up{12.1} \\
\bottomrule
\end{tabular}%
}
\end{table*}

Table~\ref{tab:lrm_ablation} reveals that the selection of LRM exerts a great influence on final performance. Gemini-3.1-Pro as LRM yields consistent improvements across all LALMs, averaging +22\% overall, with particularly large gains +28\% for open-source LALMs. However, GPT-5.2 produces sharply asymmetric effects. It substantially boosts weak LALMs but consistently degrades strong proprietary LALMs, dropping by 9\% overall. 
This degradation is most pronounced on multi-hop tasks, suggesting that GPT-5.2 may introduce spurious causal cues when reasoning over captions, thereby weakening temporal grounding and accurate event tracking.
Gemini-3.1-Pro is less affected by this failure mode, likely due to its stronger reasoning ability for long-context understanding. 
These results suggest that reasoning-enhanced cascading is not a plug-and-play solution for long-form audio understanding.
Its effectiveness depends critically on whether the external LRM can faithfully aggregate audio evidence over long caption sequences.
For weaker LALMs, a strong LRM can act as an external memory and reasoning module, compensating for limited native long-context aggregation.
However, for stronger proprietary LALMs, an ill-matched LRM may become a new bottleneck, introducing extra hallucinations.
This indicates that future cascaded systems should not simply route clip captions into generic reasoning models. Instead, they need audio-grounded reasoning mechanisms that can re-listen the original audio to verify audio evidence.

\subsection{Paralinguistic Understanding Analysis}
\label{sec:analysis_perception}

\begin{figure}[t]
  \centering
\includegraphics[width=0.999\linewidth]{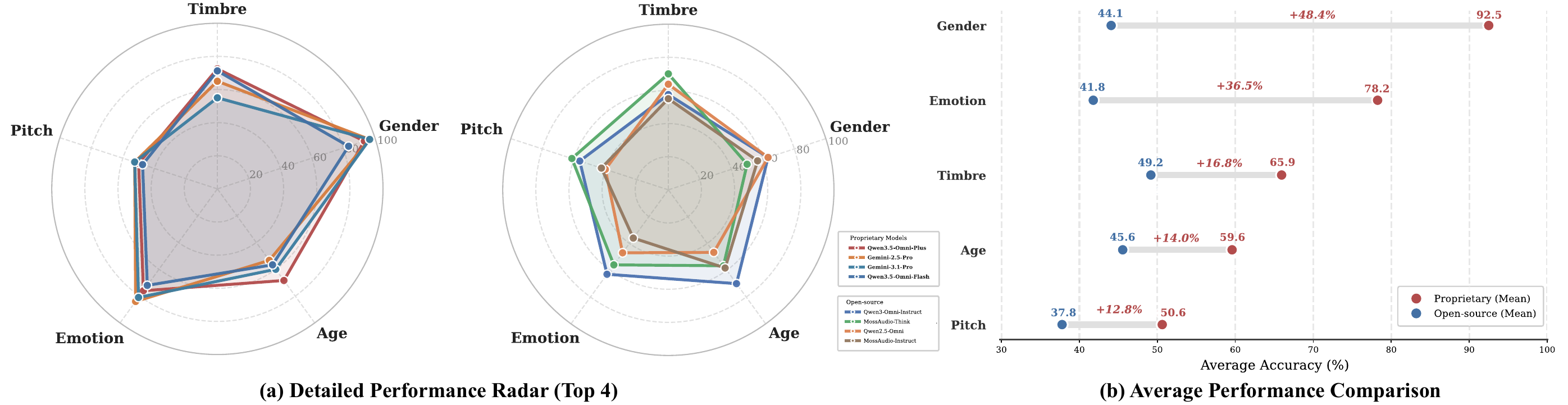}
    % \fbox{\rule[-.5cm]{0cm}{3.0cm} \rule[-.5cm]{4.2cm}{0cm}}
  \caption{Paralinguistic understanding across five fine-grained attributes (\textit{Gender}, \textit{Age}, \textit{Emotion}, \textit{Pitch}, and \textit{Timbre}) for leading proprietary versus open-source models in the cascaded caption inference setting. }
  \label{fig:para}
  \vspace{-0.cm}
\end{figure}

Figure~\ref{fig:para} reveals a pronounced performance gap between proprietary and open-source models across paralinguistic dimensions. Proprietary models achieve substantially higher accuracy on coarse-grained demographic and affective perception, reaching 92.5\% on Gender and 78.2\% on Emotion on average.
In contrast, open-source models lag behind most severely on Gender prediction, with a 48.4\% gap, indicating a clear weakness in modeling basic speaker characteristics. 
However, this gap is not uniform across all attributes.
Open-source models are relatively more competitive on Age and Timbre recognition, suggesting that some salient acoustic cues can still be captured reasonably well.

Most notably, Pitch perception emerges as a shared failure mode for both model families, with the lowest average accuracy among all paralinguistic attributes for open-source models (37.8\%) and proprietary models (50.6\%).
This indicates that current LALMs remain limited in modeling fine-grained frequency dynamics, even when they can capture semantic or coarse acoustic cues.
Overall, these results highlight the need for future LALMs to better understand paralinguistic domains. 

% \vspace{-2mm}

\begin{table*}[t]
\centering
\small
\setlength{\tabcolsep}{3pt}
\caption{Task-level accuracy (\%) on Chinese vs.\ English. $\Delta = \mathrm{Acc}_{\mathrm{EN}} - \mathrm{Acc}_{\mathrm{ZH}}$. Negative $\Delta$ indicates Chinese-favored performance and positive $\Delta$ indicates English-favored. \textit{Overall} aggregates all six task categories at the sample level (i.e., $\sum$correct/$\sum$total).}
\label{tab:lang_6task}
\resizebox{\textwidth}{!}{
\begin{tabular}{l ccc ccc ccc ccc ccc ccc ccc}
\toprule
\multirow{2}{*}{\textbf{Model}} & \multicolumn{3}{c}{\textbf{Temp. Loc.}} & \multicolumn{3}{c}{\textbf{Sem. Con.}} & \multicolumn{3}{c}{\textbf{Aco. Evt.}} & \multicolumn{3}{c}{\textbf{Paralin.}} & \multicolumn{3}{c}{\textbf{Causal}} & \multicolumn{3}{c}{\textbf{Evt. Trk.}} & \multicolumn{3}{c}{\textbf{Overall}} \\
 & ZH & EN & $\Delta$ & ZH & EN & $\Delta$ & ZH & EN & $\Delta$ & ZH & EN & $\Delta$ & ZH & EN & $\Delta$ & ZH & EN & $\Delta$ & ZH & EN & $\Delta$ \\
\midrule
\multicolumn{22}{l}{\textit{LALMs from Chinese}} \\
Qwen3.5-Omni-Plus & 51.2 & 38.4 & -12.8 & 94.0 & 90.0 & -4.0 & 64.0 & 57.0 & -7.0 & 72.6 & 66.3 & -6.3 & 76.8 & 76.0 & -0.8 & 56.8 & 56.8 & +0.0 & 68.8 & 63.6 & -5.2 \\
Qwen3-Omni-Instruct & 30.4 & 31.2 & +0.8 & 75.0 & 63.0 & -12.0 & 52.0 & 45.0 & -7.0 & 61.7 & 61.1 & -0.6 & 48.0 & 43.2 & -4.8 & 51.2 & 43.2 & -8.0 & 52.9 & 48.3 & -4.7 \\
Qwen2.5-Omni & 39.2 & 36.8 & -2.4 & 49.0 & 47.0 & -2.0 & 34.0 & 29.0 & -5.0 & 54.3 & 46.9 & -7.4 & 42.4 & 44.0 & +1.6 & 31.2 & 30.4 & -0.8 & 42.5 & 39.6 & -2.9 \\
MOSSAudio-Instruct & 28.0 & 29.6 & +1.6 & 43.0 & 38.0 & -5.0 & 43.0 & 35.0 & -8.0 & 46.9 & 48.0 & +1.1 & 38.4 & 32.0 & -6.4 & 25.6 & 26.4 & +0.8 & 37.7 & 35.6 & -2.1 \\
\midrule
\multicolumn{22}{l}{\textit{LALMs from American}} \\
Gemini-3.1-Pro & 81.6 & 77.6 & -4.0 & 96.0 & 99.0 & +3.0 & 63.0 & 51.0 & -12.0 & 69.1 & 63.4 & -5.7 & 84.0 & 86.4 & +2.4 & 68.8 & 70.4 & +1.6 & 76.4 & 73.9 & -2.5 \\
Gemini-2.5-Pro & 79.2 & 66.4 & -12.8 & 94.0 & 97.0 & +3.0 & 60.0 & 56.0 & -4.0 & 69.1 & 67.4 & -1.7 & 74.4 & 74.4 & +0.0 & 68.0 & 72.0 & +4.0 & 73.6 & 71.6 & -2.0 \\
Audio-Flamingo-3 & 34.4 & 36.8 & +2.4 & 40.0 & 47.0 & +7.0 & 44.0 & 37.0 & -7.0 & 36.6 & 48.0 & +11.4 & 61.6 & 67.2 & +5.6 & 37.6 & 43.2 & +5.6 & 42.0 & 46.9 & +4.9 \\
Phi-4-Multimodal & 30.4 & 30.4 & +0.0 & 28.0 & 32.0 & +4.0 & 27.0 & 36.0 & +9.0 & 27.4 & 40.0 & +12.6 & 31.2 & 28.8 & -2.4 & 22.4 & 24.8 & +2.4 & 27.7 & 32.4 & +4.7 \\
\bottomrule
\end{tabular}
}
\end{table*}
\subsection{Language Bias Analysis}
\label{sec:analysis_duration}
To systematically evaluate cross-lingual generalization, we analyze the performance disparity between English and Chinese inputs, defined as $\Delta = \mathrm{Acc}_{\mathrm{EN}} - \mathrm{Acc}_{\mathrm{ZH}}$.
Table~\ref{tab:lang_6task} shows that language bias is generally moderate at the overall level, but exhibits clear model- and task-dependent patterns.

First, Chinese-origin LALMs consistently perform better on Chinese inputs.
All four Chinesemodels obtain negative overall $\Delta$ values, ranging from $-2.1$\% to $-5.2$\%, indicating a stable Chinese-favored tendency,
In contrast, U.S.-origin models show a more heterogeneous pattern.
The two Gemini models also slightly favor Chinese, whereas others favor English.
This suggests that language bias is not determined solely by model origin, but is also affected by multilingual training coverage and audio-language alignment quality. 
At the task level, the largest cross-lingual discrepancies appear in tasks that depend on paralinguistic cues.
For Chinese-origin models, the Chinese advantage is particularly visible in Semantic Content, Acoustic Events, and Paralinguistic understanding.
For U.S.-origin open-source models, the opposite tendency is more pronounced in Paralinguistic understanding: Audio-Flamingo-3 and Phi-4-Multimodal show positive paralinguistic gaps of $+11.4$\% and $+12.6$\%, respectively.

These results reveal that cross-lingual robustness in long-form audio understanding is not merely a text-level issue. Even tasks that should rely heavily on acoustic evidence exhibit substantial language-dependent variation.
This suggests that current LALMs do not learn fully language-agnostic acoustic representations. 
Overall, these findings highlight the need for more balanced multilingual audio data and stronger acoustic-language disentanglement to improve cross-lingual generalization.

% \begin{figure}[hbpt]
%   \centering
%   \begin{subfigure}[b]{0.49\textwidth}
%     \centering
%     \includegraphics[width=0.99\linewidth]{Figures/fig3_avg_per_qa_grouped.pdf}
%     % \fbox{\rule[-.5cm]{0cm}{3.0cm} \rule[-.5cm]{4.2cm}{0cm}}
%     \caption{Multilingual Robustness across Task Type}
%     \label{fig:fig3_avg_per_qa_grouped}
%   \end{subfigure}\hfill
%   \begin{subfigure}[b]{0.49\textwidth}
%     \centering
%     \includegraphics[width=0.99\linewidth]{Figures/lang.pdf}
%     % \fbox{\rule[-.5cm]{0cm}{3.0cm} \rule[-.5cm]{4.2cm}{0cm}}
%     \caption{Robustness across Models by Geographic Origin}
%     \label{fig:lang}
%   \end{subfigure}\hfill
%   % \vspace{-0.15cm}
%   \caption{ (a) Investigates the cross-lingual performance across tasks. (b) Explores the multilingual capabilities of different models categorized by geographic origin across diverse language benchmarks.}
%   \label{fig:combined_stat}
%   \vspace{-0.cm}
% \end{figure}

% \input{Tables/error_breakdown}

\section{Conclusion}
\label{sec:conclusion}

We introduce \benchname, a bilingual benchmark for evaluating hour-scale audio understanding in LALMs across real-world scenarios with interleaved speech, sound, and music.
Through a systematic evaluation of broad LALMs, we show that this setting remains far from solved. Most models fall below the passing threshold, with major weaknesses in paralinguistic perception and multi-hop reasoning that require perceptual fidelity and persistent memory.
Our analyses further reveal key bottlenecks, including duration-induced performance degradation, model-dependent inference paradigms, limited long-form memory persistence, language bias, and weakness in pitch perception.
These findings suggest that future LALMs require stronger long-context audio understanding and reasoning, and more robust memory mechanisms.
We release the benchmark to support future research on hour-scale audio understanding.

\newpage

% =======================================================
%                     References
% =======================================================
\bibliographystyle{plainnat}

\end{CJK*}
\end{document}